\documentclass[twocolumn,secnumarabic,amssymb, amsmath, nobibnotes, aps, prl, floatfix]{revtex4-2}

\usepackage{graphicx}
\usepackage[dvipsnames]{xcolor}
\usepackage{cleveref}
\usepackage{etoolbox}
\usepackage{tikz}
\usetikzlibrary{arrows.meta}
\usetikzlibrary{shapes.geometric}
\usepackage{pgfplots}
\usepackage{marginnote} 
\usepackage[inline]{enumitem}

\begin{document}

\newcommand{\GB}[1]{[{\color{orange}#1}]}	
\newcommand{\LB}[1]{[{\color{magenta}#1}]}
\newcommand{\SC}[1]{[{\color{red}\textbf{#1}}]}
\newcommand{\MC}[1]{[{\color{blue}#1}]}
\newcommand{\SM}[1]{[{\color{orange}#1}]}
\newcommand{\wel}{$We_{\mathcal{L}}$}
\newcommand{\mur}{$\mu_d/\mu_c$}
\newcommand{\lra}[1]{\langle #1 \rangle }
\definecolor{lightblue}{RGB}{173,216,230} 

\title{How small droplets form in turbulent multiphase flows 
}%

\author{M. Crialesi-Esposito$^{1}$}
\author{G. Boffetta$^{2}$}
\author{L. Brandt$^{3,4}$}
\author{S. Chibbaro$^{5}$}
\author{S. Musacchio$^{2}$
}

\affiliation{
${1}$ DIEF, University of Modena and Reggio Emilia, 41125 Modena, Italy\\
${2}$ Dipartimento di Fisica and INFN, Universit\`a degli Studi di Torino, via P. Giuria 1, 10125 Torino, Italy.\\
$3$ FLOW Centre, KTH Royal Institute of Technology, Stockholm, Sweden\\
$4$ Department of Energy and Process Engineering, Norwegian University of Science and Technology(NTNU), Trondheim, Norway\\
$5$ Universit\'e Paris-Saclay, CNRS, LISN, 91400 Orsay, France
}

\begin{abstract}
The formation of small droplets and bubbles in turbulent flows is a crucial
process in geophysics and engineering, whose underlying physical mechanism
remains a puzzle.  In this letter, we address this problem by means of
high-resolution numerical simulations, comparing a realistic multiphase
configuration with a 
numerical experiment
in which we attenuate the
presence of strong velocity gradients either across the whole mixture or 
in the disperse phase only. 
Our results show unambiguously that the formation of
small droplets is governed by the internal dynamics which occurs during the
break-up of large drops and that the high vorticity and the extreme dissipation associated to these
events are the consequence and not the cause of the breakup.

\end{abstract}

\maketitle
\paragraph{Introduction}

The dynamics of droplet and bubble breakup in turbulence is fundamental for
several industrial \cite{mcclements2015food} and environmental processes
\cite{Deane2002,Li1998}.  Because of the complex turbulent environment, drops
typically have a broad range of sizes.  In several cases, the diameter of the
smallest droplets/bubbles is of the utmost importance, as for the dissolution
of air bubbles in the oceans \cite{Deike2022} or the transport of oil droplets
deep into the marine environment after spilling \cite{Li1998}.  The main actors
at play in such processes are turbulence and capillarity, with the balance
between the two determining the  minimum droplet diameter for breakup to
occur, before capillarity can resist the turbulent pressure fluctuations
causing fragmentation.  This threshold size is called the Kolmogorov-Hinze (KH)
scale~\cite{kolmogorov1949,Hinze1955}, and on dimensional considerations reads
as:
\begin{equation}
d_{KH} \sim (\rho_c/\sigma)^{-3/5}\langle\varepsilon\rangle^{-2/5},
\label{eq:hinze}
\end{equation} 
where $\rho_c$ is the density of the carrier phase, $\sigma$ is the surface
tension and $\langle \varepsilon\rangle$ is the domain averaged turbulent
energy dissipation rate.
The fragmentation dynamics for droplets larger than the KH scale is understood
in terms of a local cascade {\it à la Kolmogorov}~\cite{Garrett2000}, for which
experimental and numerical evidences have been
presented~\cite{Deane2002,riviere2021sub,crialesi2022}.
A broad spectrum of sub-Hinze droplets with diameter smaller than $d_{KH}$ is
also found. 
Despite recent attempts to understand this
regime~\cite{riviere2022capillary,qi2022fragmentation,crialesi2023interaction,vela2022memoryless},
the origin/dynamics of these small droplets in turbulence and their interaction
with the surrounding flow remain mostly unknown.

A key feature of the breakup process appears to be the presence of strong
velocity gradients in proximity of regions with high interfacial curvature,
which contributes to increasing the local vorticity and creates areas of high
energy dissipation \cite{crialesi2022}. 
Two possible complementary mechanisms are thought to be at the origin 
of the sub-Hinze droplets:
(i) the presence of local events of extreme turbulence which induce a local
decrease of the KH scale and cause the droplet breakup;
(ii) capillary dynamics, which leads to a pinch-off and eventually generates
intense dissipative events.


\begin{figure}[h!]
	\includegraphics[width=0.23\textwidth]{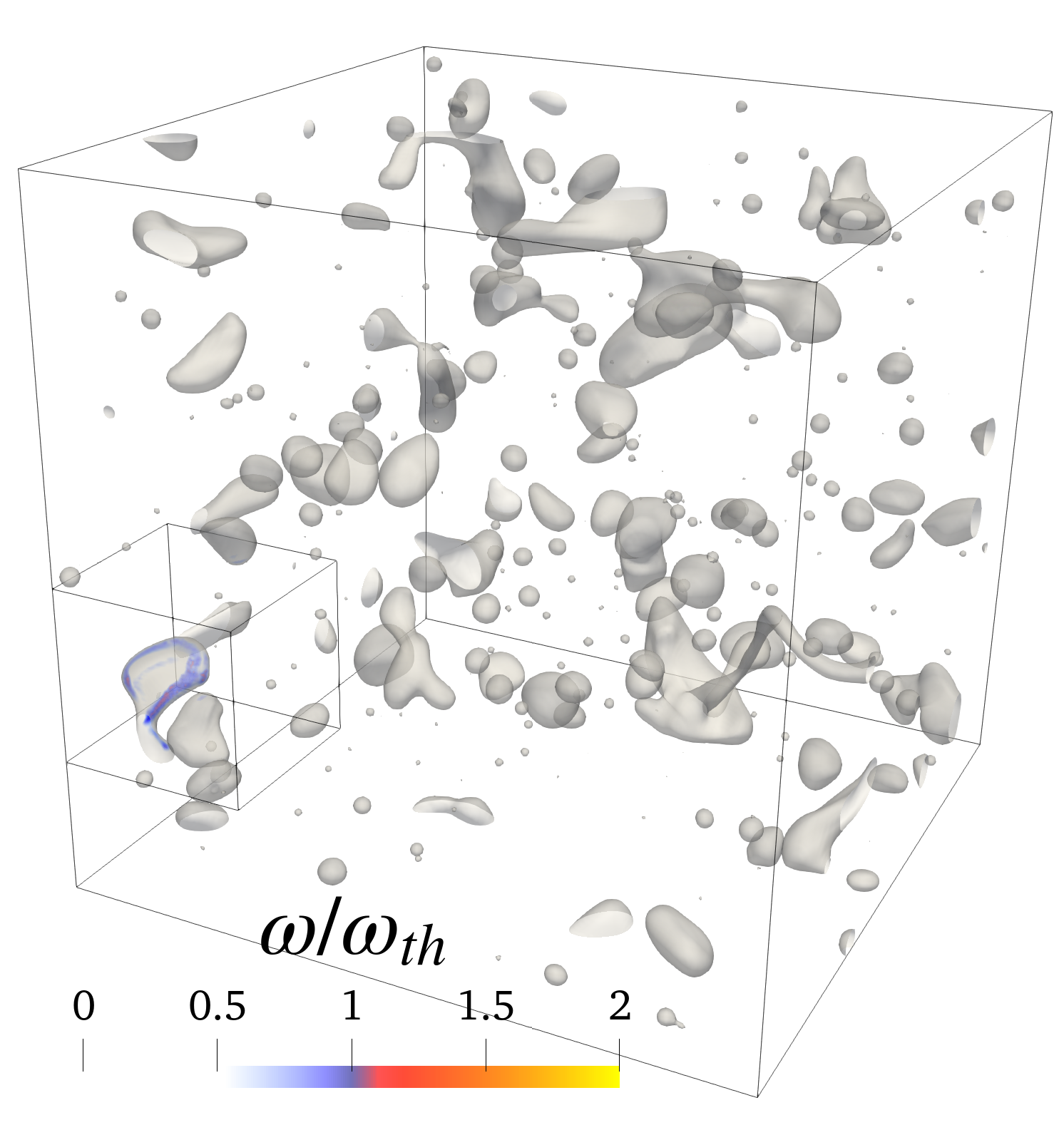}
		\put(-110,110){\large\textbf{(\textit{a})}}
	\includegraphics[width=0.23\textwidth]{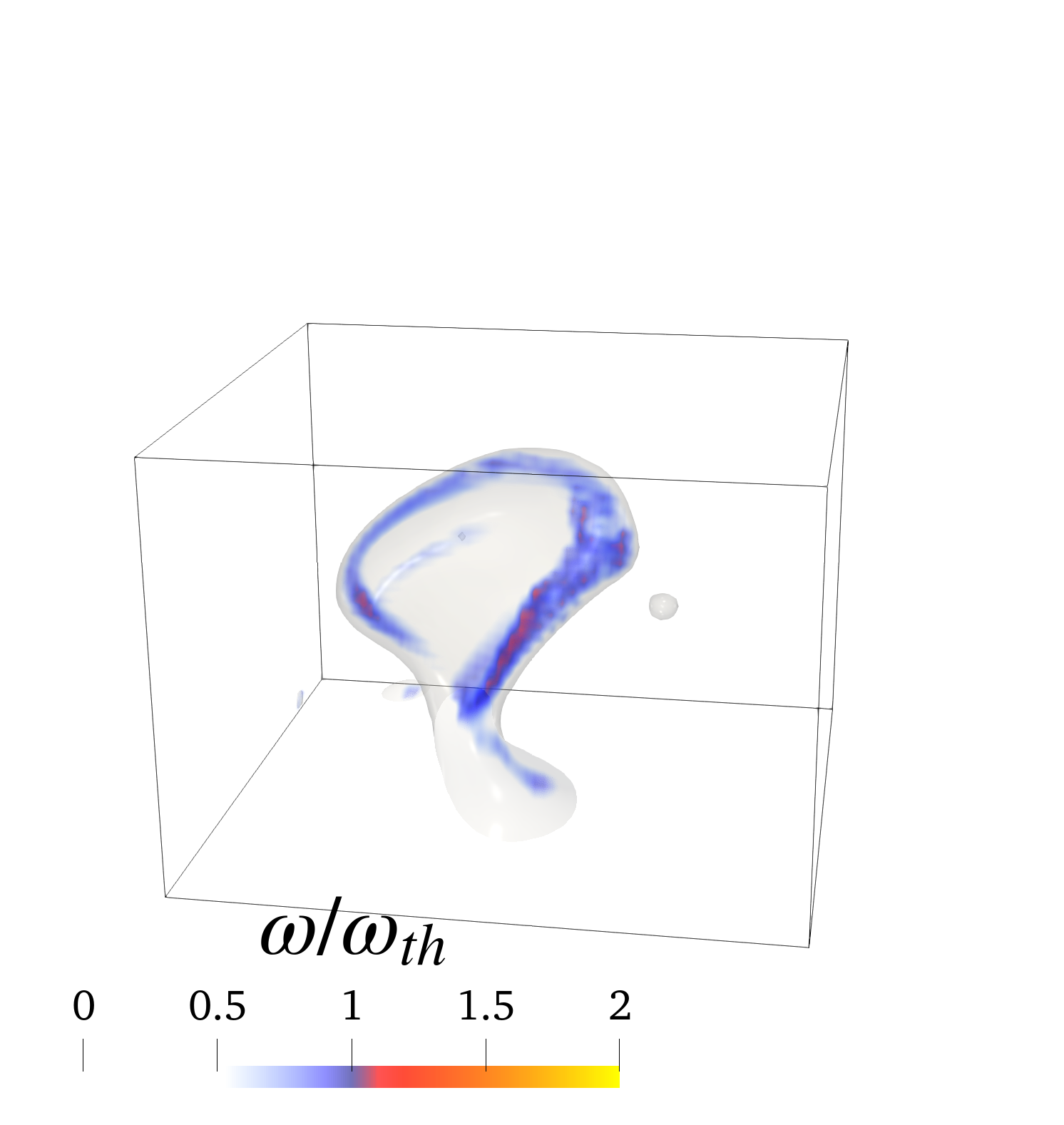}
		\put(-110,110){\large\textbf{(\textit{b})}}
	
	\includegraphics[width=0.23\textwidth]{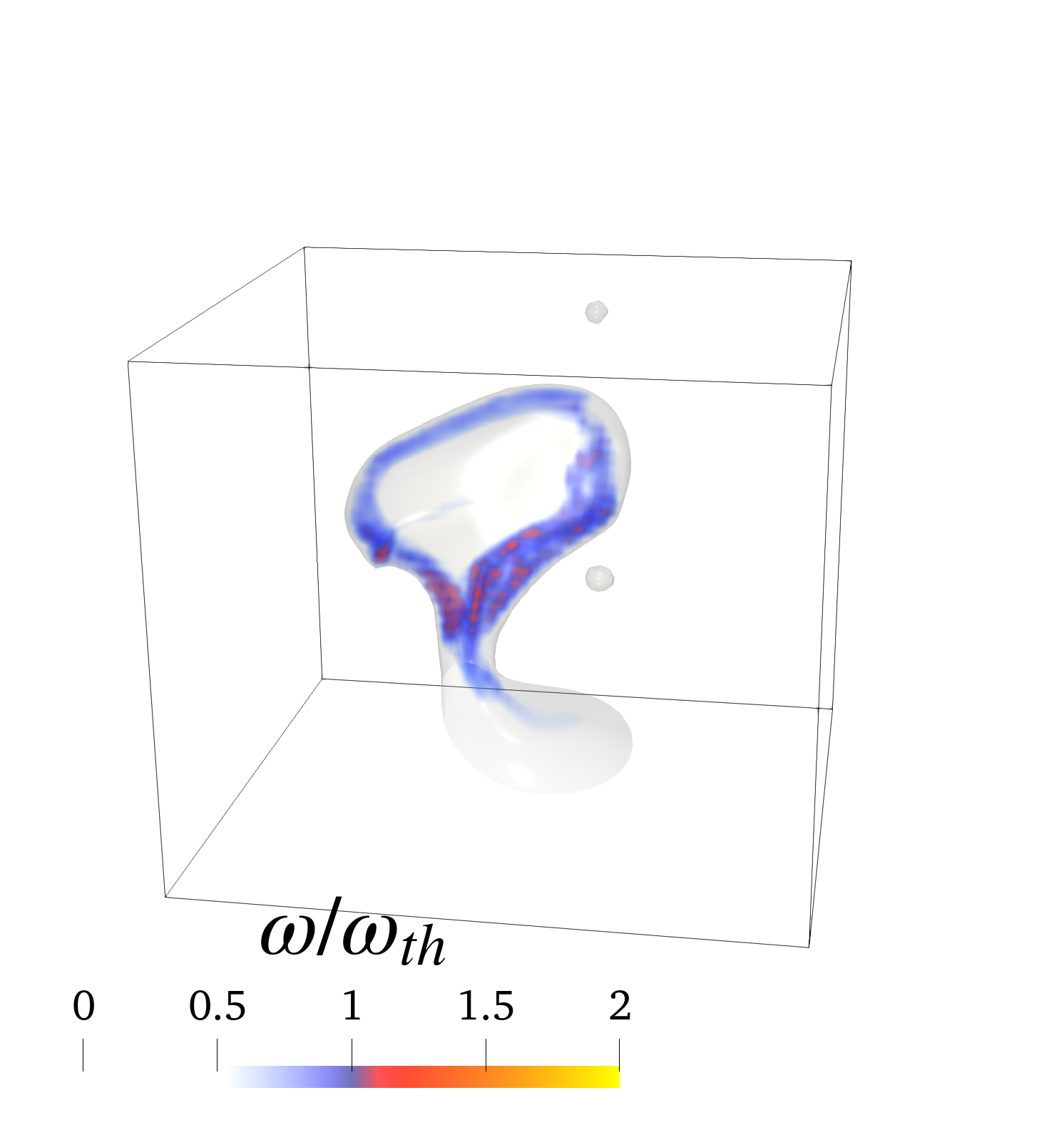}
	\put(-110,110){\large\textbf{(\textit{c})}}
	\includegraphics[width=0.23\textwidth]{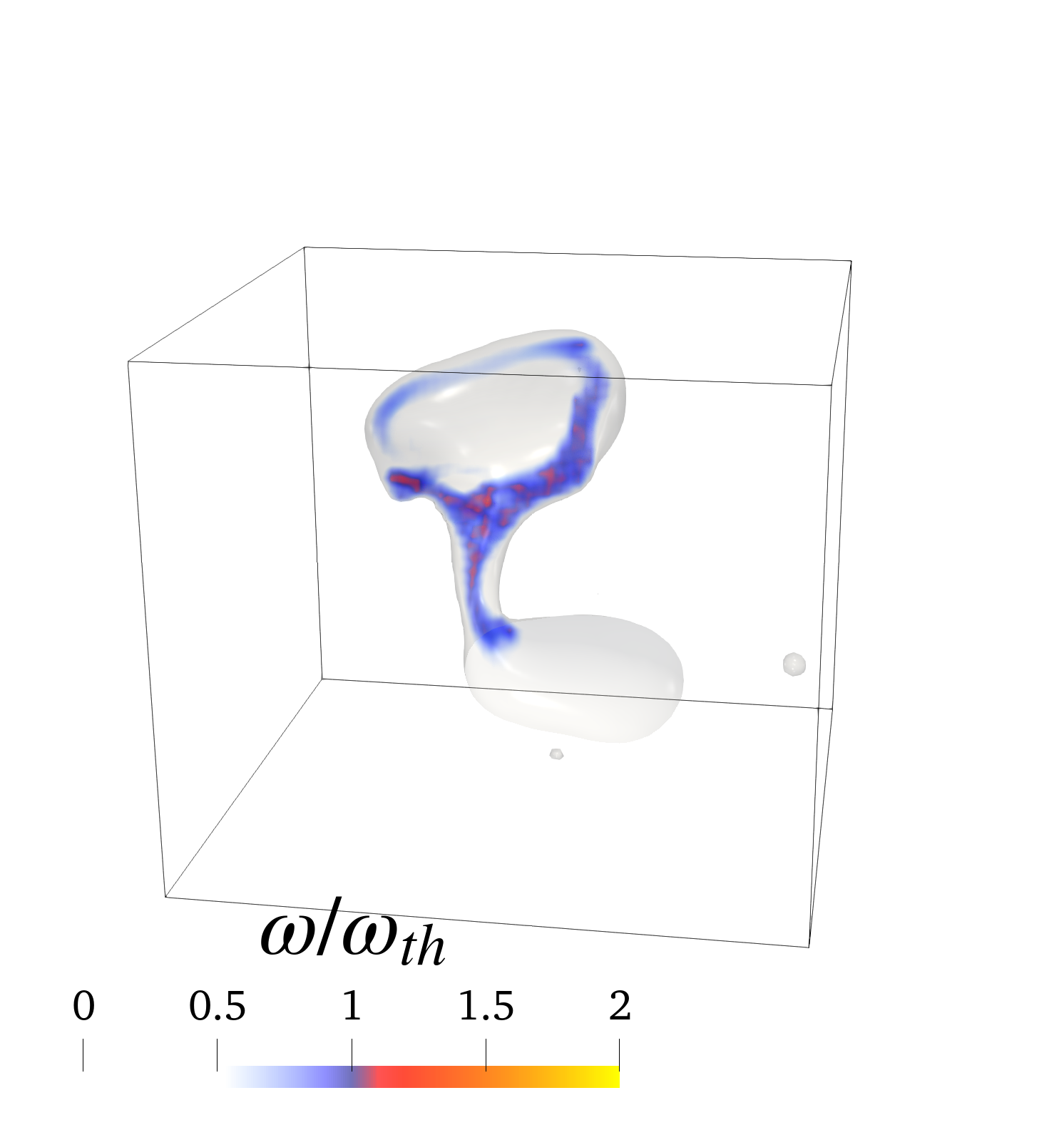}
	\put(-110,110){\large\textbf{(\textit{d})}}
	
	\includegraphics[width=0.23\textwidth]{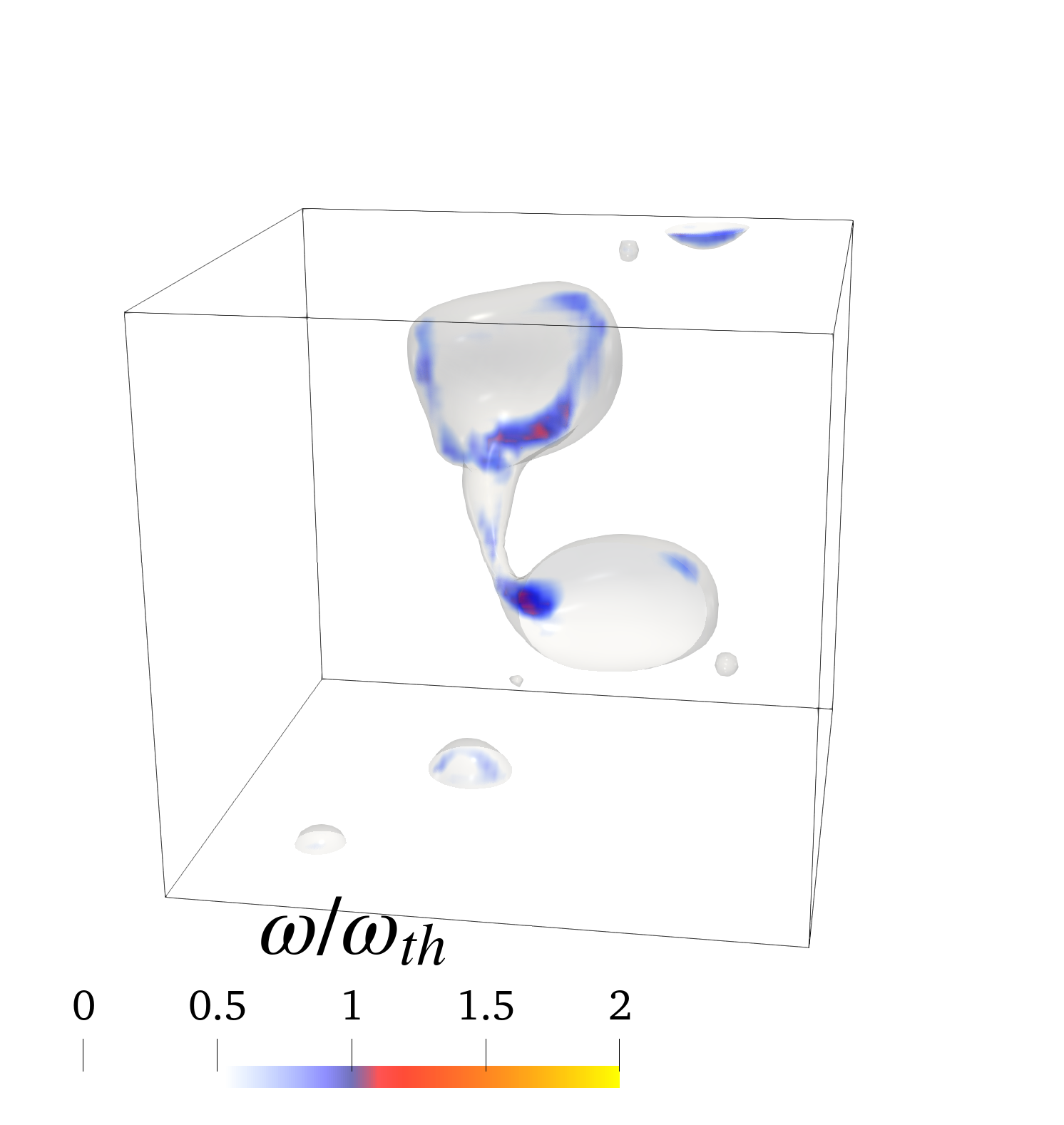}
	\put(-110,110){\large\textbf{(\textit{e})}}
	\includegraphics[width=0.23\textwidth]{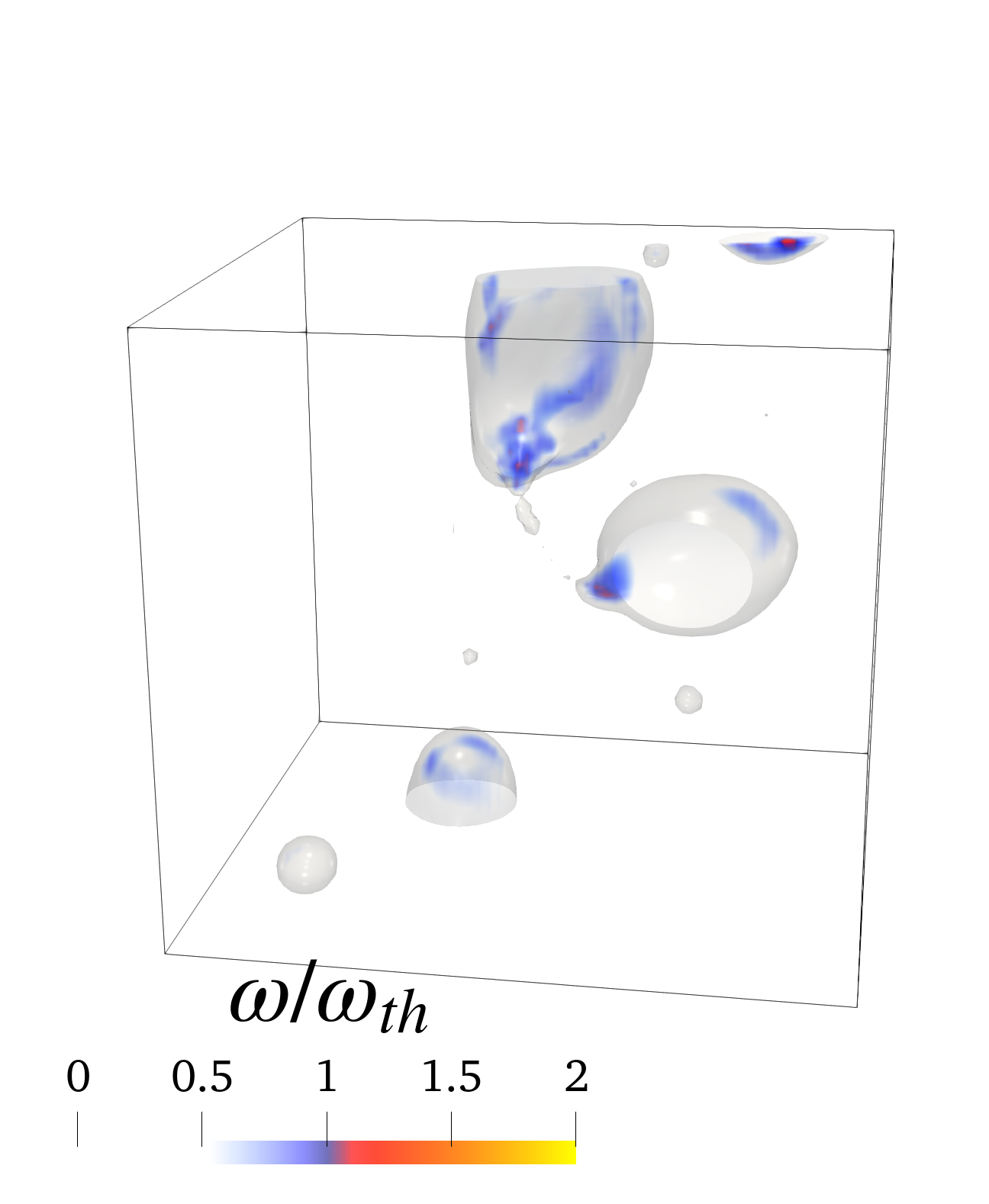}
	\put(-110,110){\large\textbf{(\textit{f})}}

\caption{Visualisation of high-intensity vorticity within the droplet during
breakup. (a) a portion of the domain is extracted from the whole simulation.
(b) beginning of a droplet breakup event: the droplet separates into two
regions, divided by a neck; the vorticity inside the droplet shows high values
($\omega\approx\omega_{th}$) on the edges. (c-d) As the droplet deforms,
regions of stronger vorticity form in proximity of the neck. (e-f)  the breakup
becomes  inevitable and  smaller droplets form as a result.}
\label{fig1}
\end{figure}

In the favour of the second scenario, experiments on the fragmentation of 
a single drop have shown that the sub-Hinze dynamics is non-local
in size~\cite{qi2022fragmentation,crialesi2023interaction},
suggesting that the turbulence strain produces small filaments, which
eventually leads to pinch-off via fast capillary dynamics~\cite{ruth2019bubble}.
One numerical example is shown in Figure~\ref{fig1}, where a droplet 
larger than the KH size is deformed by turbulence into a thin ligament which,
by Rayleigh-Plateau instabilities, produces several small droplets. Non 
locality is evident since the size of the daughter droplet is related with the diameter of 
the ligament and not of the entire droplet.
Within this scenario, large dissipation events are induced by the
rupture of the interface at the origin of sub-Hinze structures.

On the other hand, recent works have pointed out that the presence of the
interface increases the probability of large vorticity and
dissipative regions \cite{fuster2021vortex}, leading to a higher intermittency  than
single-phase turbulence~\cite{crialesi2022,crialesi2023interaction}. This
evidence might support the first mechanism (i), meaning that intense vorticity
external to the droplets is the main cause of the fragmentation and the
formation of sub-Hinze inclusions.
Thus far, no numerical or laboratory experiment has been able to settle this
issue and the statistical relevance of the two possible scenarios 
in particular in a realistic configuration with many droplets.


In this Letter, we address this dilemma by means of high-resolution 
numerical simulations, in which we control the small scale dynamics by 
penalizing the vorticity field, something not possible in a laboratory 
experiment \cite{Buzzicotti2020}.
The idea comes directly from~\Cref{eq:hinze}.
In a local sense, $d\sim\varepsilon^{-2/5}$,
suggests that small droplets are linked to regions of high 
dissipation.
Hence, artificially penalizing the regions with strong velocity gradients
(where $\varepsilon\gg \langle\varepsilon\rangle$)
should enable us to understand where and how sub-Hinze structures are formed.
Thanks to this surgery of the turbulent flow, we are capable to 
show that it is the flow {\it inside} the droplets which dominates the 
formation of very small droplets and the dynamics at the smallest scales,
therefore supporting the second of the proposed mechanisms. 
Moreover, we show that  turbulent extreme events have some impact on the
structures with a diameter of the order of the KH scale, but are statistically
negligible for the scales in the sub-Hinze range, where capillary effects
are dominant.  Our results demonstrate that the formation of small droplets
below the KH scale is primarily dominated by capillarity, and, in turn, that
the generation of small droplets is responsible for the presence of local
maxima of vorticity and dissipation, rather than the contrary.

\paragraph{Methods}
We solve the one-fluid Navier-Stokes equations (NSE) including deformable
interfaces~\cite{tryggvason2011direct}:
\begin{equation}
 \frac{d \mathbf{u}}{d t}
 = -{\nabla P} + 
 \nabla \cdot \left[\nu([\nabla \mathbf{u} +\nabla \mathbf{u}^{\text{T}})\right]
  + \sigma\xi\delta_s \mathbf{n}  + \mathbf{f} + \mathbf{f^C} ,
	\label{eq:ns}
\end{equation}
where $\mathbf{u}$ is the velocity field, $P$ is the pressure, $\nu$ is the kinematic viscosity, $\xi$ is the interface curvature, $\mathbf{n}$ is the interface normal, $\delta_s$ is a delta-Dirac function localised at the interface between the two phases, and $\sigma$ is the surface tension coefficient. Turbulence is sustained through the ABC forcing $\mathbf{f}$~\cite{Mininni2006}.

To control the flow we add the penalizing term $\mathbf{f}^C=-C \mathbf{u}$ ~\cite{Buzzicotti2020},
where regions of high vorticity $\omega$ can be suppressed directly in the momentum equation through the masking function:
\begin{equation}
	C = \beta\left(\frac{tanh(\omega-\omega_{th})+1}{2}\right)~,
	\label{eq:filt}
\end{equation}
where $\beta$ is the filter amplitude ($\beta=0$ corresponds to standard NSE),
$\omega=|\sum_{ij}(\partial_iu_j-\partial_j u_i)|$ is the vorticity modulus,
and $\omega_{th}=5\sigma_{\omega}$ is the maximum threshold value,
with $\sigma_\omega$ the standard deviation of the vorticity for the reference multiphase case (see below). 

It is worth  noticing that a penalization force which suppresses the regions
with large values of the energy dissipation rate $\varepsilon$
cannot be applied because its effect would be canceled by the pressure
gradients.
In a preliminary test, we observed that a direct masking based on $\varepsilon$
alters the local structure of the velocity-gradient tensor and it does not
preserve the incompressibility of the velocity field.  Enforcing the
incompressibility restores the local velocity gradients and cancels the
effect of the penalization force.       To overcome this issue we adopt a
penalization method that suppresses the regions of high vorticity, which are
linked to the events of strong dissipation while having a different local
structure of the velocity gradients~\cite{Frisch1995a}.

Simulations are carried out {\it via} the open-source code FLUTAS~\cite{crialesi2023flutas}, and the interface is reconstructed using the Volume of Fluid method MTHINC~\cite{Ii2012}.
Simulations are performed at the Taylor-scale Reynolds number $Re_\lambda=137$, measured in the single-phase turbulent field~\cite{crialesi2022}.
The box-side length is $2\pi$, discretised using $N=512$ grid-points, with turbulence sustained at $L_f=\pi$, with a kinematic viscosity $\nu=0.006$, and a matching density and viscosity between the phases. 
The volume fraction is $\alpha=V_d/V=0.1$, where $V_d$ is the volume of the dispersed phase and $V$ is the volume of the computational domain.
The large-scale Weber number is $We=\rho u_{rms}^2 L_f/\sigma=42.6$.
The present setup has been shown to develop a droplet distribution $N(d)$
displaying both the $-10/3$ and $-3/2$ scaling ranges at scales larger
and smaller than the KH scale~\cite{crialesi2022}. 

\paragraph{Numerical Results}
We report results from four numerical experiments, comparing three 
different multiphase simulations and one single-phase.  
Reference simulations are for single-phase (SP) and multiphase (MP) flow.
The two simulations in which the vorticity is penalized 
in the whole domain (MPp) and only inside the dispersed phase (MPp,i). 
For the results corresponding to the  single-phase penalized simulation see Supplemental Material.
\begin{figure}[h]
\includegraphics[width=0.3\textwidth]{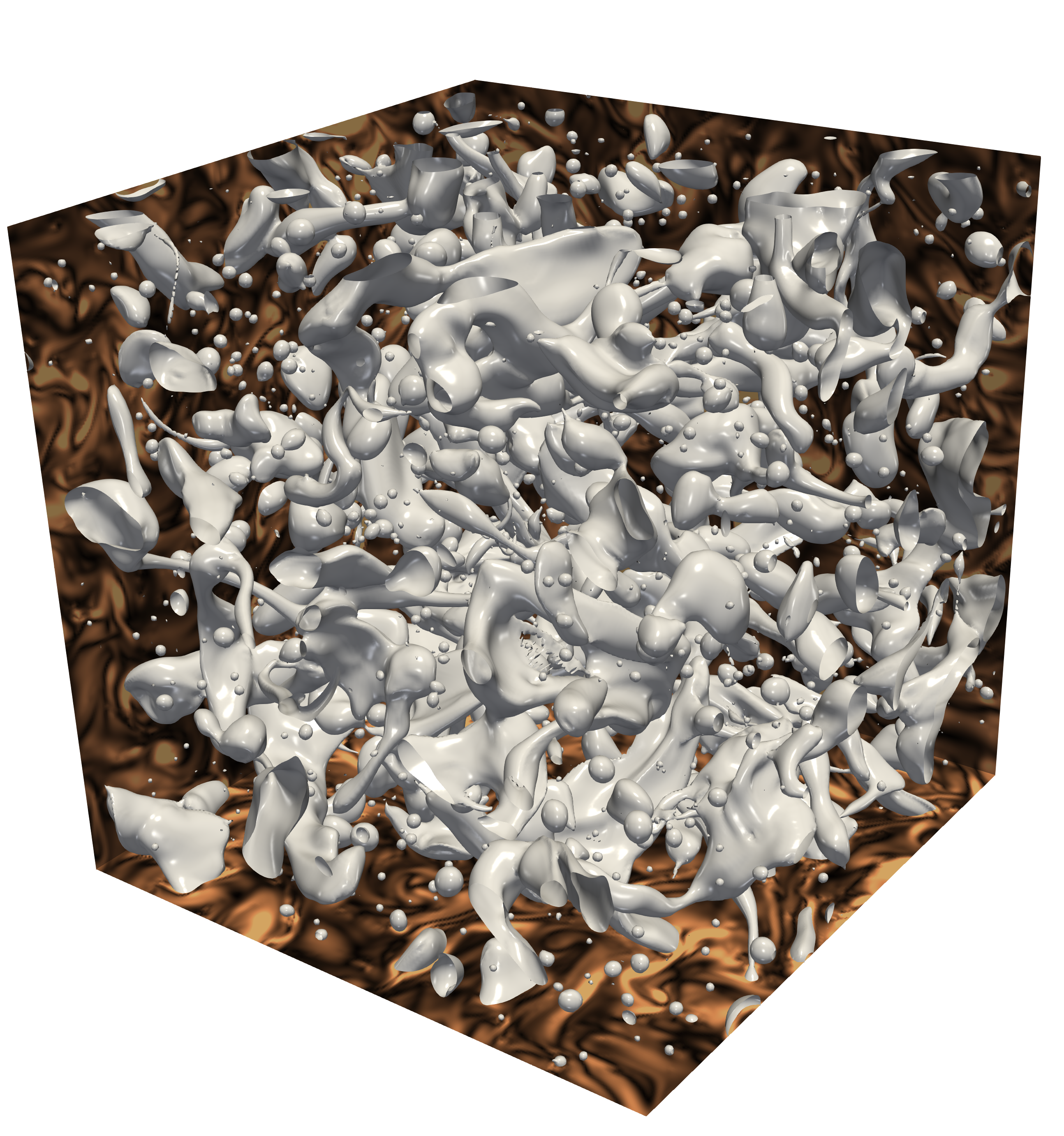}
\put(-190,150){\large\textbf{(\textit{a})}}

\includegraphics[width=0.3\textwidth]{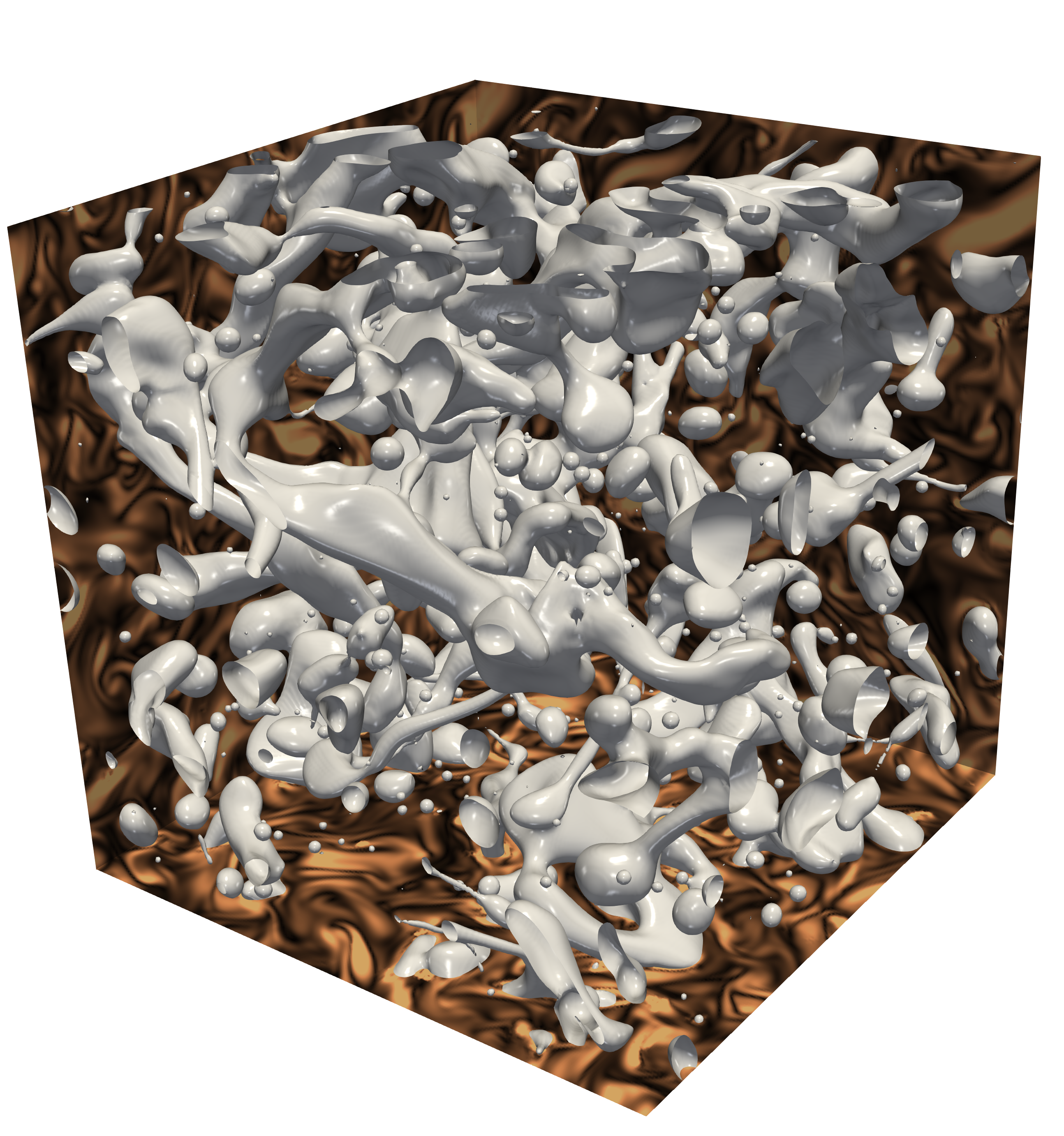}
\put(-190,150){\large\textbf{(\textit{b})}}
\caption{Render of (\textit{a}) non filtered MP flow and (\textit{b}) filtered
MPp flow.  We show iso-contours of the color function (the droplet interface)
and the vorticity field projected on the background planes.}
\label{fig2}
\end{figure}

From the visual comparison of the MP and MPp simulations shown in
Figure~\ref{fig2}, we may already qualitatively observe  that suppressing local
gradients partially inhibits the formation of small droplets, while preserving
the large-scale flow structures.  In particular, in the multiphase simulation
with the penalization force (MP, \Cref{fig2}(\textit{b}) we observe the
formation of elongated fluid structures stretched by large-scale vortices, but
the fragmentation of the droplets is overall attenuated.  

\begin{figure}[h]
\includegraphics[width=0.4\textwidth]{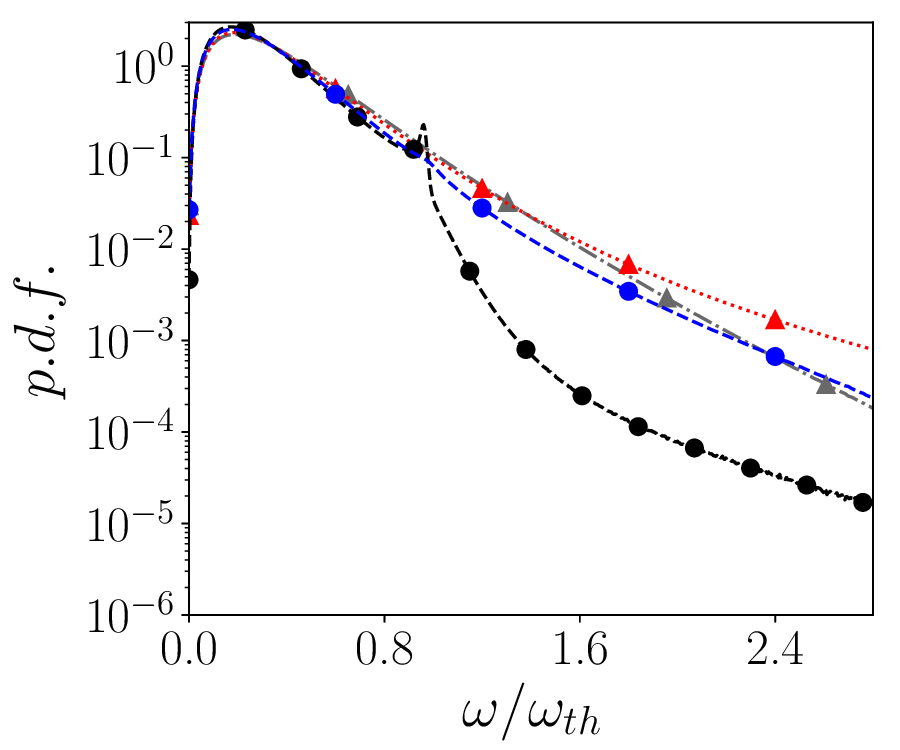}
\put(-190,180){\large{\textbf{(\textit{a})}}}
	
\includegraphics[width=0.4\textwidth]{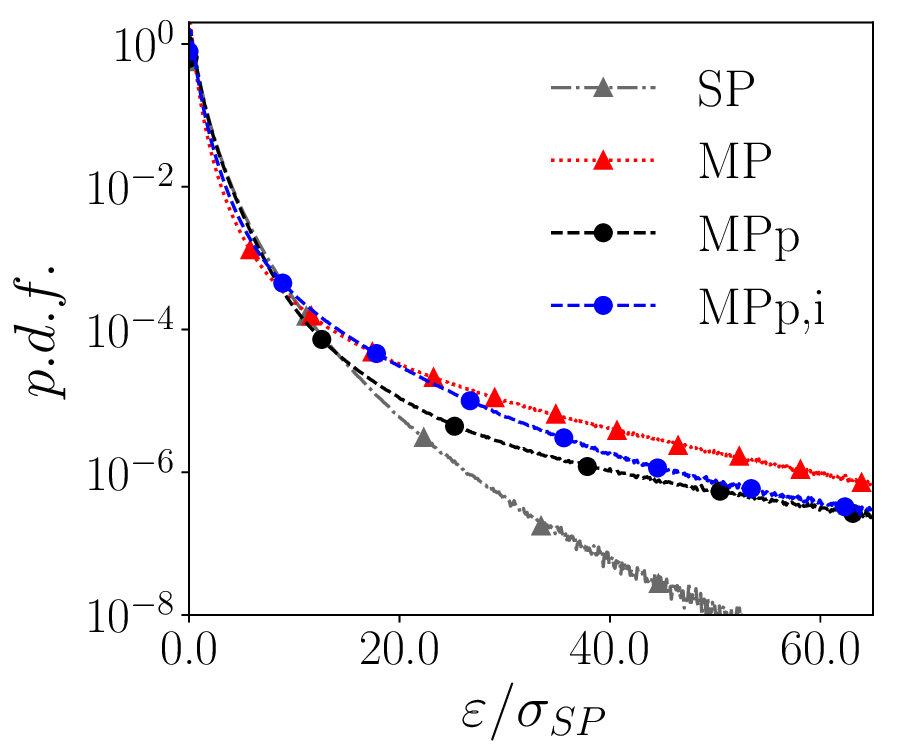}
\put(-190,180){\large\textbf{(\textit{b})}}
\caption{Probability density functions of vorticity magnitude, panel (\textit{a}) and energy dissipation $\varepsilon\equiv\nu|\partial_i u_j|^2$, panel (\textit{b}).  
 SP and MP refer to $C=0$ cases, while MPp and MPp,i are the penalized  cases, for which $\omega_{th}=5\sigma_\omega$ ($\sigma_\omega$ being the vorticity standard deviation for case MP) and $\beta=0.02$.
 Vorticity is normalised by $\omega_{th}$,  while energy dissipation by its
standard deviation for the single-phase case $\sigma_{SP}$. The label $i$
stands for inside, which indicates when the vorticity is penalised only
inside the droplet phase, which corresponds to the $\alpha=10\%$ of the
mixture.
}
\label{fig3}
\end{figure}

To quantify the penalization effects, we report in \Cref{fig3} the probability
density function (PDF) of the vorticity magnitude (panel (\textit{a})) and of
the energy dissipation  (panel (\textit{b})).  Comparing with the SP case,
the MP flow displays an increment of the PDF tails  both for vorticity and
dissipation, confirming that the presence of the interface increases
intermittency~\cite{crialesi2023interaction}. The
increase is substantial for the energy dissipation.  
The effect of the filter appears in the vorticity PDFs sharply at 
$\omega=\omega_{th}$, i.e.\ exactly at
the masking threshold. High values of $\omega$ are still possible due
to the incompressibility constraint of \Cref{eq:ns}, as observed also
for Navier-Stokes, SP simulations~\cite{Buzzicotti2020}.
From Figure~\ref{fig3}a it is evident that the statistics of vorticity
obtained penalising inside the droplet phase only (MPp,i) is  close to the unfiltered field MP, because of the
penalization only acts on 10\% of the total volume.
The PDF of dissipation appears to be less affected by penalization.
We find that the average dissipation in the MPp run is reduced of 
about $15\%$ with respect to the MP case. According to 
\Cref{eq:hinze} this correspond to a change of $6\%$ in $d_{KH}$.

\begin{figure}
\includegraphics[width=0.475\textwidth]{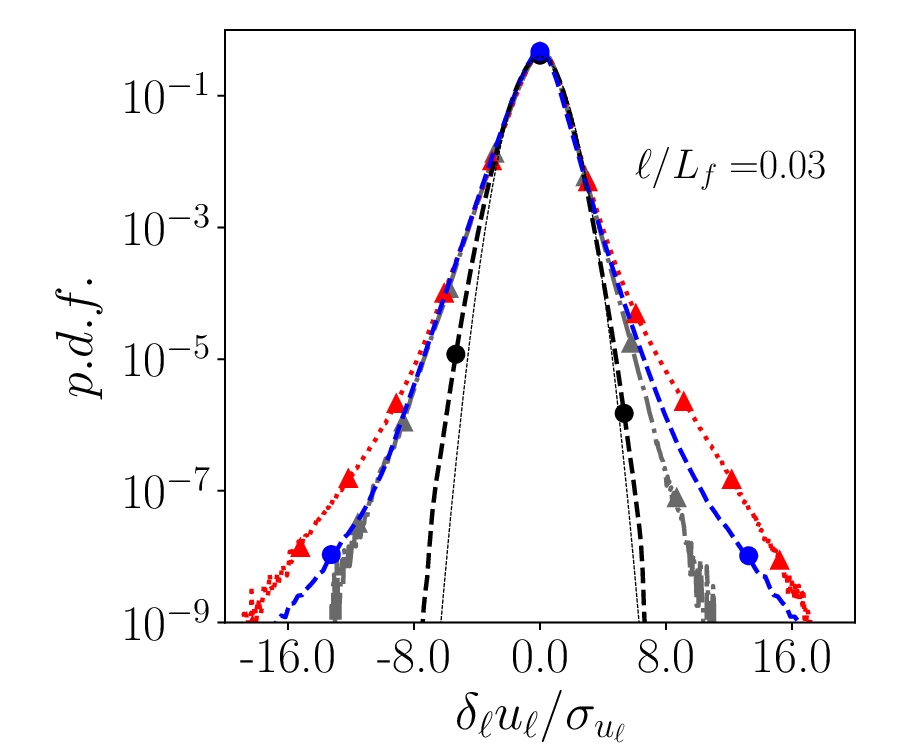}
\caption{PDF of velocity increments at $\ell=0.03L_f$, normalized by the standard deviation for case SP.  The PDF is obtained at small scales, much
smaller than the Kolmogorov-Hinze scale. The legend is the same as in Figure
~\ref{fig3}.}
\label{fig4}
\end{figure}

Further insight on the flow statistics is provided by the analysis of the PDF of the velocity
increments $\delta_\ell u_\ell=u(x+\ell)-u(x)$~\cite{monin2013statistical},
reported in \Cref{fig4} for a separation below the Hinze scale.
In the absence of penalization, the multiphase case (MP) is more intermittent,
as also recently investigated~\cite{crialesi2023intermittency}.
When the mask is acting on the whole fluid (MPp case)
the distributions become gaussian-like in multiphase flows,
similarly to the single-phase case discussed in Ref.~\cite{Buzzicotti2020}.  
However, if the vorticity is filtered only inside the dispersed phase (MPp,i case) 
the probability distributions vary little when compared to the SP flow,
the curves being distinguishable only in the far tails.
This is explained by the low volume fraction considered, as most of the field (i.e.\ 90\%) remains unchanged. 
The data lie between those of the the single phase and the multiphase flow, since the masking is not removing 
 the  interfaces, which are responsible for the increase in intermittency.
Which scales are affected by the masking function is discussed in the Supplemental Material, 
where we show that penalizing strong vorticity regions is equivalent to act at scales below the Kolmogorov-Hinze one.


\begin{figure}
	\includegraphics[width=0.47\textwidth]{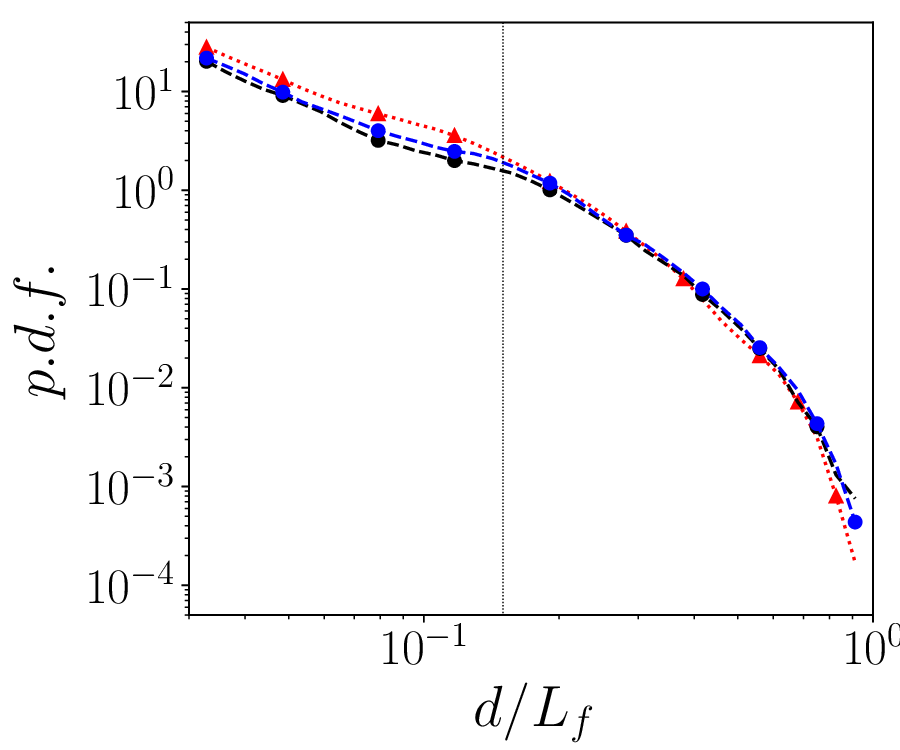}
	\put(-185,37){\includegraphics[width=0.20\textwidth]{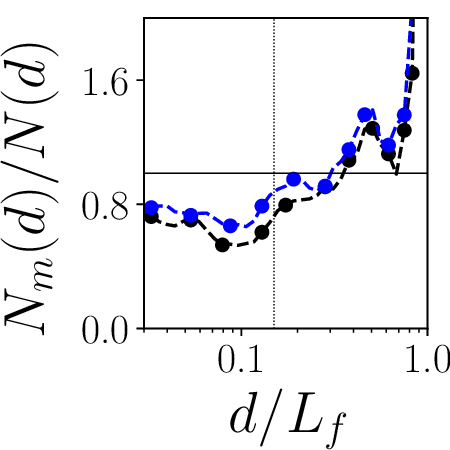}}
	\caption{
	Droplet-size-distribution for cases MP (red dotted line with triangles),  MPp (black dashed line with dots), and MPp,i (blue dashed line with dots) where probability is obtained by normalizing all cases for the total number of droplets of the MP case. The inset shows the ratio for the total number of droplets of the filtered cases and the reference MP case. 
   The black dotted vertical line represents the KH scale
	}
	\label{fig5}
\end{figure}

The effect of vorticity penalization (both in the dispersed phase and in the whole flow) on the droplet-size distribution is shown in \Cref{fig5} together with the comparison with the unmasked case. 
To ease the comparison, we normalize the different curves 
by the total number of droplets of the unmasked MP flow.
The distribution of large droplets (above the KH scale) for both the MPp and MPp,i cases are close to the MP reference run. 
However, significant quantitative differences are found in the total number of small droplets, especially for $d<d_{KH}\approx 0.15L_f$; notably, the number of droplets in the MPp flow is approximately 60\% of those in the MP case. 
This can be better appreciated in the inset of \Cref{fig5}, where we display
the ratio between the number of droplets of the masked cases and the unmasked
MP case.
As the mass is conserved, the number of larger droplets increases in the
masked cases.
The most important result is that the distribution is the same when the mask
is applied to the whole mixture or only inside the dispersed phase.
This observation is at odd with 
the statistics of 
dissipation which are almost unaffected by the penalization in the 
dispersed phase and demonstrates that high levels of vorticity and the most extreme dissipation events are a
consequence of the destabilization process which leads to the breakup.

\paragraph{Discussion}

The present results indicate that the mechanism underlying the production of droplets below the Kolmogorov-Hinze scale in a turbulent emulsion is reduced when the vorticity is limited inside the droplets. Consequently, the origin of very small droplets can be traced back to capillary stresses, which act faster than the smallest turbulent eddies.

The following dynamical process has been unveiled: turbulent motions deform the droplets locally creating filaments, typically larger than the Kolmogorov-Hinze scale. Capillary instabilies are then triggered by the turbulent fluctuations (appearance of necks). This further reduces the filament dimension and produces vorticity inside the droplet, which, in turn, accelerates the filament instability in a self-sustained process, eventually leading to the rupture into different drops.

This last singular step is associated with strong vorticity release and increased dissipation. Therefore, extreme dissipation events are not the cause, but rather the effect of the break-up.  Note also that the physics of the fragmentation of droplets larger than $d_{KH}$ is instead largely unaffected by the masking, confirming the quasi-local cascade {\it à la Kolmogorov} in this range.

Our findings show that it is possible to simplify the study of small droplets formation  in turbulence by neglecting the action of large scale motion, and focusing on the droplet deformed state and its internal dynamics.
For future works, it would be precious to analyse in detail the dynamics of the rupture of ligaments in relation with the creation of vorticity and dissipation. That would be important to build up reduced models relevant for applications.
\paragraph{Acknowledgments}
M.C.E. acknowledges the financial support given by the Department of Engineering ‘Enzo
Ferrari’ of the University of Modena and Reggio Emilia through the action ‘FAR dipartimentale 2023/2024’. S.M. and M.C.E. acknowledge EuroHPC for awarding access to MeluXina GPU through the project EHPC-REG-2022R01-052. 
\bibliography{references_new}

\begin{thebibliography}{23}%
\makeatletter
\providecommand \@ifxundefined [1]{%
 \@ifx{#1\undefined}
}%
\providecommand \@ifnum [1]{%
 \ifnum #1\expandafter \@firstoftwo
 \else \expandafter \@secondoftwo
 \fi
}%
\providecommand \@ifx [1]{%
 \ifx #1\expandafter \@firstoftwo
 \else \expandafter \@secondoftwo
 \fi
}%
\providecommand \natexlab [1]{#1}%
\providecommand \enquote  [1]{``#1''}%
\providecommand \bibnamefont  [1]{#1}%
\providecommand \bibfnamefont [1]{#1}%
\providecommand \citenamefont [1]{#1}%
\providecommand \href@noop [0]{\@secondoftwo}%
\providecommand \href [0]{\begingroup \@sanitize@url \@href}%
\providecommand \@href[1]{\@@startlink{#1}\@@href}%
\providecommand \@@href[1]{\endgroup#1\@@endlink}%
\providecommand \@sanitize@url [0]{\catcode `\\12\catcode `\$12\catcode
  `\&12\catcode `\#12\catcode `\^12\catcode `\_12\catcode `\%12\relax}%
\providecommand \@@startlink[1]{}%
\providecommand \@@endlink[0]{}%
\providecommand \url  [0]{\begingroup\@sanitize@url \@url }%
\providecommand \@url [1]{\endgroup\@href {#1}{\urlprefix }}%
\providecommand \urlprefix  [0]{URL }%
\providecommand \Eprint [0]{\href }%
\providecommand \doibase [0]{https://doi.org/}%
\providecommand \selectlanguage [0]{\@gobble}%
\providecommand \bibinfo  [0]{\@secondoftwo}%
\providecommand \bibfield  [0]{\@secondoftwo}%
\providecommand \translation [1]{[#1]}%
\providecommand \BibitemOpen [0]{}%
\providecommand \bibitemStop [0]{}%
\providecommand \bibitemNoStop [0]{.\EOS\space}%
\providecommand \EOS [0]{\spacefactor3000\relax}%
\providecommand \BibitemShut  [1]{\csname bibitem#1\endcsname}%
\let\auto@bib@innerbib\@empty
\bibitem [{\citenamefont {McClements}(2015)}]{mcclements2015food}%
  \BibitemOpen
  \bibfield  {author} {\bibinfo {author} {\bibfnamefont {D.~J.}\ \bibnamefont
  {McClements}},\ }\href@noop {} {\emph {\bibinfo {title} {{Food emulsions:
  principles, practices, and techniques}}}}\ (\bibinfo  {publisher} {CRC
  press},\ \bibinfo {year} {2015})\BibitemShut {NoStop}%
\bibitem [{\citenamefont {Deane}\ and\ \citenamefont
  {Stokes}(2002)}]{Deane2002}%
  \BibitemOpen
  \bibfield  {author} {\bibinfo {author} {\bibfnamefont {G.~B.}\ \bibnamefont
  {Deane}}\ and\ \bibinfo {author} {\bibfnamefont {M.~D.}\ \bibnamefont
  {Stokes}},\ }\bibfield  {title} {\bibinfo {title} {{Scale dependence of
  bubble creation mechanisms in breaking waves}},\ }\href
  {https://doi.org/10.1038/nature00967} {\bibfield  {journal} {\bibinfo
  {journal} {Nature}\ }\textbf {\bibinfo {volume} {418}},\ \bibinfo {pages}
  {839} (\bibinfo {year} {2002})}\BibitemShut {NoStop}%
\bibitem [{\citenamefont {Li}\ and\ \citenamefont {Garrett}(1998)}]{Li1998}%
  \BibitemOpen
  \bibfield  {author} {\bibinfo {author} {\bibfnamefont {M.}~\bibnamefont
  {Li}}\ and\ \bibinfo {author} {\bibfnamefont {C.}~\bibnamefont {Garrett}},\
  }\bibfield  {title} {\bibinfo {title} {{The relationship between oil droplet
  size and upper ocean turbulence}},\ }\href
  {https://doi.org/10.1016/S0025-326X(98)00096-4} {\bibfield  {journal}
  {\bibinfo  {journal} {Marine Pollution Bulletin}\ }\textbf {\bibinfo {volume}
  {36}},\ \bibinfo {pages} {961} (\bibinfo {year} {1998})}\BibitemShut
  {NoStop}%
\bibitem [{\citenamefont {Deike}(2022)}]{Deike2022}%
  \BibitemOpen
  \bibfield  {author} {\bibinfo {author} {\bibfnamefont {L.}~\bibnamefont
  {Deike}},\ }\bibfield  {title} {\bibinfo {title} {{Mass Transfer at the
  Ocean–Atmosphere Interface: The Role of Wave Breaking, Droplets, and
  Bubbles}},\ }\href {https://doi.org/10.1146/annurev-fluid-030121-014132}
  {\bibfield  {journal} {\bibinfo  {journal} {Annual Review of Fluid
  Mechanics}\ }\textbf {\bibinfo {volume} {54}},\ \bibinfo {pages} {191}
  (\bibinfo {year} {2022})}\BibitemShut {NoStop}%
\bibitem [{\citenamefont {Kolmogorov}(1949)}]{kolmogorov1949}%
  \BibitemOpen
  \bibfield  {author} {\bibinfo {author} {\bibfnamefont {A.}~\bibnamefont
  {Kolmogorov}},\ }\bibfield  {title} {\bibinfo {title} {{On the breakage of
  drops in a turbulent flow}},\ }\href@noop {} {\bibfield  {journal} {\bibinfo
  {journal} {Dokl. Akad. Navk. SSSR}\ }\textbf {\bibinfo {volume} {66}},\
  \bibinfo {pages} {825} (\bibinfo {year} {1949})}\BibitemShut {NoStop}%
\bibitem [{\citenamefont {Hinze}(1955)}]{Hinze1955}%
  \BibitemOpen
  \bibfield  {author} {\bibinfo {author} {\bibfnamefont {J.~O.}\ \bibnamefont
  {Hinze}},\ }\bibfield  {title} {\bibinfo {title} {{Fundamentals of the
  hydrodynamic mechanism of splitting in dispersion processes}},\ }\href
  {https://doi.org/10.1002/aic.690010303} {\bibfield  {journal} {\bibinfo
  {journal} {AIChE Journal}\ }\textbf {\bibinfo {volume} {1}},\ \bibinfo
  {pages} {289} (\bibinfo {year} {1955})}\BibitemShut {NoStop}%
\bibitem [{\citenamefont {Garrett}\ \emph {et~al.}(2000)\citenamefont
  {Garrett}, \citenamefont {Li},\ and\ \citenamefont {Farmer}}]{Garrett2000}%
  \BibitemOpen
  \bibfield  {author} {\bibinfo {author} {\bibfnamefont {C.}~\bibnamefont
  {Garrett}}, \bibinfo {author} {\bibfnamefont {M.}~\bibnamefont {Li}},\ and\
  \bibinfo {author} {\bibfnamefont {D.}~\bibnamefont {Farmer}},\ }\bibfield
  {title} {\bibinfo {title} {{The connection between bubble size spectra and
  energy dissipation rates in the upper ocean}},\ }\href
  {https://doi.org/10.1175/1520-0485(2000)030<2163:TCBBSS>2.0.CO;2} {\bibfield
  {journal} {\bibinfo  {journal} {Journal of Physical Oceanography}\ }\textbf
  {\bibinfo {volume} {30}},\ \bibinfo {pages} {2163} (\bibinfo {year}
  {2000})}\BibitemShut {NoStop}%
\bibitem [{\citenamefont {Rivi{\`e}re}\ \emph {et~al.}(2021)\citenamefont
  {Rivi{\`e}re}, \citenamefont {Mostert}, \citenamefont {Perrard},\ and\
  \citenamefont {Deike}}]{riviere2021sub}%
  \BibitemOpen
  \bibfield  {author} {\bibinfo {author} {\bibfnamefont {A.}~\bibnamefont
  {Rivi{\`e}re}}, \bibinfo {author} {\bibfnamefont {W.}~\bibnamefont
  {Mostert}}, \bibinfo {author} {\bibfnamefont {S.}~\bibnamefont {Perrard}},\
  and\ \bibinfo {author} {\bibfnamefont {L.}~\bibnamefont {Deike}},\ }\bibfield
   {title} {\bibinfo {title} {Sub-hinze scale bubble production in turbulent
  bubble break-up},\ }\href@noop {} {\bibfield  {journal} {\bibinfo  {journal}
  {Journal of Fluid Mechanics}\ }\textbf {\bibinfo {volume} {917}},\ \bibinfo
  {pages} {A40} (\bibinfo {year} {2021})}\BibitemShut {NoStop}%
\bibitem [{\citenamefont {Crialesi-Esposito}\ \emph {et~al.}(2022)\citenamefont
  {Crialesi-Esposito}, \citenamefont {Rosti}, \citenamefont {Chibbaro},\ and\
  \citenamefont {Brandt}}]{crialesi2022}%
  \BibitemOpen
  \bibfield  {author} {\bibinfo {author} {\bibfnamefont {M.}~\bibnamefont
  {Crialesi-Esposito}}, \bibinfo {author} {\bibfnamefont {M.~E.}\ \bibnamefont
  {Rosti}}, \bibinfo {author} {\bibfnamefont {S.}~\bibnamefont {Chibbaro}},\
  and\ \bibinfo {author} {\bibfnamefont {L.}~\bibnamefont {Brandt}},\
  }\bibfield  {title} {\bibinfo {title} {Modulation of homogeneous and
  isotropic turbulence in emulsions},\ }\href
  {https://doi.org/10.1017/jfm.2022.179} {\bibfield  {journal} {\bibinfo
  {journal} {Journal of Fluid Mechanics}\ }\textbf {\bibinfo {volume} {940}},\
  \bibinfo {pages} {A19} (\bibinfo {year} {2022})}\BibitemShut {NoStop}%
\bibitem [{\citenamefont {Rivi{\`e}re}\ \emph {et~al.}(2022)\citenamefont
  {Rivi{\`e}re}, \citenamefont {Ruth}, \citenamefont {Mostert}, \citenamefont
  {Deike},\ and\ \citenamefont {Perrard}}]{riviere2022capillary}%
  \BibitemOpen
  \bibfield  {author} {\bibinfo {author} {\bibfnamefont {A.}~\bibnamefont
  {Rivi{\`e}re}}, \bibinfo {author} {\bibfnamefont {D.~J.}\ \bibnamefont
  {Ruth}}, \bibinfo {author} {\bibfnamefont {W.}~\bibnamefont {Mostert}},
  \bibinfo {author} {\bibfnamefont {L.}~\bibnamefont {Deike}},\ and\ \bibinfo
  {author} {\bibfnamefont {S.}~\bibnamefont {Perrard}},\ }\bibfield  {title}
  {\bibinfo {title} {Capillary driven fragmentation of large gas bubbles in
  turbulence},\ }\href@noop {} {\bibfield  {journal} {\bibinfo  {journal}
  {Physical Review Fluids}\ }\textbf {\bibinfo {volume} {7}},\ \bibinfo {pages}
  {083602} (\bibinfo {year} {2022})}\BibitemShut {NoStop}%
\bibitem [{\citenamefont {Qi}\ \emph {et~al.}(2022)\citenamefont {Qi},
  \citenamefont {Tan}, \citenamefont {Corbitt}, \citenamefont {Urbanik},
  \citenamefont {Salibindla},\ and\ \citenamefont {Ni}}]{qi2022fragmentation}%
  \BibitemOpen
  \bibfield  {author} {\bibinfo {author} {\bibfnamefont {Y.}~\bibnamefont
  {Qi}}, \bibinfo {author} {\bibfnamefont {S.}~\bibnamefont {Tan}}, \bibinfo
  {author} {\bibfnamefont {N.}~\bibnamefont {Corbitt}}, \bibinfo {author}
  {\bibfnamefont {C.}~\bibnamefont {Urbanik}}, \bibinfo {author} {\bibfnamefont
  {A.~K.}\ \bibnamefont {Salibindla}},\ and\ \bibinfo {author} {\bibfnamefont
  {R.}~\bibnamefont {Ni}},\ }\bibfield  {title} {\bibinfo {title}
  {Fragmentation in turbulence by small eddies},\ }\href@noop {} {\bibfield
  {journal} {\bibinfo  {journal} {Nature Communications}\ }\textbf {\bibinfo
  {volume} {13}},\ \bibinfo {pages} {1} (\bibinfo {year} {2022})}\BibitemShut
  {NoStop}%
\bibitem [{\citenamefont {Crialesi-Esposito}\ \emph
  {et~al.}(2023{\natexlab{a}})\citenamefont {Crialesi-Esposito}, \citenamefont
  {Chibbaro},\ and\ \citenamefont {Brandt}}]{crialesi2023interaction}%
  \BibitemOpen
  \bibfield  {author} {\bibinfo {author} {\bibfnamefont {M.}~\bibnamefont
  {Crialesi-Esposito}}, \bibinfo {author} {\bibfnamefont {S.}~\bibnamefont
  {Chibbaro}},\ and\ \bibinfo {author} {\bibfnamefont {L.}~\bibnamefont
  {Brandt}},\ }\bibfield  {title} {\bibinfo {title} {The interaction of droplet
  dynamics and turbulence cascade},\ }\href@noop {} {\bibfield  {journal}
  {\bibinfo  {journal} {Communications Physics}\ }\textbf {\bibinfo {volume}
  {6}},\ \bibinfo {pages} {5} (\bibinfo {year}
  {2023}{\natexlab{a}})}\BibitemShut {NoStop}%
\bibitem [{\citenamefont {Vela-Mart{\'\i}n}\ and\ \citenamefont
  {Avila}(2022)}]{vela2022memoryless}%
  \BibitemOpen
  \bibfield  {author} {\bibinfo {author} {\bibfnamefont {A.}~\bibnamefont
  {Vela-Mart{\'\i}n}}\ and\ \bibinfo {author} {\bibfnamefont {M.}~\bibnamefont
  {Avila}},\ }\bibfield  {title} {\bibinfo {title} {Memoryless drop breakup in
  turbulence},\ }\href@noop {} {\bibfield  {journal} {\bibinfo  {journal}
  {Science Advances}\ }\textbf {\bibinfo {volume} {8}},\ \bibinfo {pages}
  {eabp9561} (\bibinfo {year} {2022})}\BibitemShut {NoStop}%
\bibitem [{\citenamefont {Ruth}\ \emph {et~al.}(2019)\citenamefont {Ruth},
  \citenamefont {Mostert}, \citenamefont {Perrard},\ and\ \citenamefont
  {Deike}}]{ruth2019bubble}%
  \BibitemOpen
  \bibfield  {author} {\bibinfo {author} {\bibfnamefont {D.~J.}\ \bibnamefont
  {Ruth}}, \bibinfo {author} {\bibfnamefont {W.}~\bibnamefont {Mostert}},
  \bibinfo {author} {\bibfnamefont {S.}~\bibnamefont {Perrard}},\ and\ \bibinfo
  {author} {\bibfnamefont {L.}~\bibnamefont {Deike}},\ }\bibfield  {title}
  {\bibinfo {title} {Bubble pinch-off in turbulence},\ }\href@noop {}
  {\bibfield  {journal} {\bibinfo  {journal} {Proceedings of the National
  Academy of Sciences}\ }\textbf {\bibinfo {volume} {116}},\ \bibinfo {pages}
  {25412} (\bibinfo {year} {2019})}\BibitemShut {NoStop}%
\bibitem [{\citenamefont {Fuster}\ and\ \citenamefont
  {Rossi}(2021)}]{fuster2021vortex}%
  \BibitemOpen
  \bibfield  {author} {\bibinfo {author} {\bibfnamefont {D.}~\bibnamefont
  {Fuster}}\ and\ \bibinfo {author} {\bibfnamefont {M.}~\bibnamefont {Rossi}},\
  }\bibfield  {title} {\bibinfo {title} {Vortex-interface interactions in
  two-dimensional flows},\ }\href@noop {} {\bibfield  {journal} {\bibinfo
  {journal} {International Journal of Multiphase Flow}\ }\textbf {\bibinfo
  {volume} {143}},\ \bibinfo {pages} {103757} (\bibinfo {year}
  {2021})}\BibitemShut {NoStop}%
\bibitem [{\citenamefont {Buzzicotti}\ \emph {et~al.}(2020)\citenamefont
  {Buzzicotti}, \citenamefont {Biferale},\ and\ \citenamefont
  {Toschi}}]{Buzzicotti2020}%
  \BibitemOpen
  \bibfield  {author} {\bibinfo {author} {\bibfnamefont {M.}~\bibnamefont
  {Buzzicotti}}, \bibinfo {author} {\bibfnamefont {L.}~\bibnamefont
  {Biferale}},\ and\ \bibinfo {author} {\bibfnamefont {F.}~\bibnamefont
  {Toschi}},\ }\bibfield  {title} {\bibinfo {title} {Statistical properties of
  turbulence in the presence of a smart small-scale control},\ }\href@noop {}
  {\bibfield  {journal} {\bibinfo  {journal} {Physical Review Letters}\
  }\textbf {\bibinfo {volume} {124}},\ \bibinfo {pages} {084504} (\bibinfo
  {year} {2020})}\BibitemShut {NoStop}%
\bibitem [{\citenamefont {Tryggvason}\ \emph {et~al.}(2011)\citenamefont
  {Tryggvason}, \citenamefont {Scardovelli},\ and\ \citenamefont
  {Zaleski}}]{tryggvason2011direct}%
  \BibitemOpen
  \bibfield  {author} {\bibinfo {author} {\bibfnamefont {G.}~\bibnamefont
  {Tryggvason}}, \bibinfo {author} {\bibfnamefont {R.}~\bibnamefont
  {Scardovelli}},\ and\ \bibinfo {author} {\bibfnamefont {S.}~\bibnamefont
  {Zaleski}},\ }\href@noop {} {\emph {\bibinfo {title} {Direct numerical
  simulations of gas--liquid multiphase flows}}}\ (\bibinfo  {publisher}
  {Cambridge university press},\ \bibinfo {year} {2011})\BibitemShut {NoStop}%
\bibitem [{\citenamefont {Mininni}\ \emph {et~al.}(2006)\citenamefont
  {Mininni}, \citenamefont {Alexakis},\ and\ \citenamefont
  {Pouquet}}]{Mininni2006}%
  \BibitemOpen
  \bibfield  {author} {\bibinfo {author} {\bibfnamefont {P.~D.}\ \bibnamefont
  {Mininni}}, \bibinfo {author} {\bibfnamefont {A.}~\bibnamefont {Alexakis}},\
  and\ \bibinfo {author} {\bibfnamefont {A.}~\bibnamefont {Pouquet}},\
  }\bibfield  {title} {\bibinfo {title} {{Large-scale flow effects, energy
  transfer, and self-similarity on turbulence}},\ }\href
  {https://doi.org/10.1103/PhysRevE.74.016303} {\bibfield  {journal} {\bibinfo
  {journal} {Physical Review E - Statistical, Nonlinear, and Soft Matter
  Physics}\ }\textbf {\bibinfo {volume} {74}},\ \bibinfo {pages} {1} (\bibinfo
  {year} {2006})}\BibitemShut {NoStop}%
\bibitem [{\citenamefont {Frisch}(1995)}]{Frisch1995a}%
  \BibitemOpen
  \bibfield  {author} {\bibinfo {author} {\bibfnamefont {U.}~\bibnamefont
  {Frisch}},\ }\href {https://doi.org/10.1017/CBO9781139170666} {\emph
  {\bibinfo {title} {{Turbulence}}}}\ (\bibinfo  {publisher} {Cambridge
  University Press},\ \bibinfo {year} {1995})\BibitemShut {NoStop}%
\bibitem [{\citenamefont {Crialesi-Esposito}\ \emph
  {et~al.}(2023{\natexlab{b}})\citenamefont {Crialesi-Esposito}, \citenamefont
  {Scapin}, \citenamefont {Demou}, \citenamefont {Rosti}, \citenamefont
  {Costa}, \citenamefont {Spiga},\ and\ \citenamefont
  {Brandt}}]{crialesi2023flutas}%
  \BibitemOpen
  \bibfield  {author} {\bibinfo {author} {\bibfnamefont {M.}~\bibnamefont
  {Crialesi-Esposito}}, \bibinfo {author} {\bibfnamefont {N.}~\bibnamefont
  {Scapin}}, \bibinfo {author} {\bibfnamefont {A.~D.}\ \bibnamefont {Demou}},
  \bibinfo {author} {\bibfnamefont {M.~E.}\ \bibnamefont {Rosti}}, \bibinfo
  {author} {\bibfnamefont {P.}~\bibnamefont {Costa}}, \bibinfo {author}
  {\bibfnamefont {F.}~\bibnamefont {Spiga}},\ and\ \bibinfo {author}
  {\bibfnamefont {L.}~\bibnamefont {Brandt}},\ }\bibfield  {title} {\bibinfo
  {title} {Flutas: A gpu-accelerated finite difference code for multiphase
  flows},\ }\href@noop {} {\bibfield  {journal} {\bibinfo  {journal} {Computer
  Physics Communications}\ }\textbf {\bibinfo {volume} {284}},\ \bibinfo
  {pages} {108602} (\bibinfo {year} {2023}{\natexlab{b}})}\BibitemShut
  {NoStop}%
\bibitem [{\citenamefont {Ii}\ \emph {et~al.}(2012)\citenamefont {Ii},
  \citenamefont {Sugiyama}, \citenamefont {Takeuchi}, \citenamefont {Takagi},
  \citenamefont {Matsumoto},\ and\ \citenamefont {Xiao}}]{Ii2012}%
  \BibitemOpen
  \bibfield  {author} {\bibinfo {author} {\bibfnamefont {S.}~\bibnamefont
  {Ii}}, \bibinfo {author} {\bibfnamefont {K.}~\bibnamefont {Sugiyama}},
  \bibinfo {author} {\bibfnamefont {S.}~\bibnamefont {Takeuchi}}, \bibinfo
  {author} {\bibfnamefont {S.}~\bibnamefont {Takagi}}, \bibinfo {author}
  {\bibfnamefont {Y.}~\bibnamefont {Matsumoto}},\ and\ \bibinfo {author}
  {\bibfnamefont {F.}~\bibnamefont {Xiao}},\ }\bibfield  {title} {\bibinfo
  {title} {{An interface capturing method with a continuous function: The THINC
  method with multi-dimensional reconstruction}},\ }\href
  {https://doi.org/10.1016/j.jcp.2011.11.038} {\bibfield  {journal} {\bibinfo
  {journal} {Journal of Computational Physics}\ }\textbf {\bibinfo {volume}
  {231}},\ \bibinfo {pages} {2328} (\bibinfo {year} {2012})}\BibitemShut
  {NoStop}%
\bibitem [{\citenamefont {Monin}\ and\ \citenamefont
  {Yaglom}(2013)}]{monin2013statistical}%
  \BibitemOpen
  \bibfield  {author} {\bibinfo {author} {\bibfnamefont {A.}~\bibnamefont
  {Monin}}\ and\ \bibinfo {author} {\bibfnamefont {A.}~\bibnamefont {Yaglom}},\
  }\href@noop {} {\emph {\bibinfo {title} {Statistical fluid mechanics:
  mechanics of turbulence}}}\ (\bibinfo  {publisher} {Courier Corporation},\
  \bibinfo {year} {2013})\BibitemShut {NoStop}%
\bibitem [{\citenamefont {Crialesi-Esposito}\ \emph
  {et~al.}(2023{\natexlab{c}})\citenamefont {Crialesi-Esposito}, \citenamefont
  {Boffetta}, \citenamefont {Brandt}, \citenamefont {Chibbaro},\ and\
  \citenamefont {Musacchio}}]{crialesi2023intermittency}%
  \BibitemOpen
  \bibfield  {author} {\bibinfo {author} {\bibfnamefont {M.}~\bibnamefont
  {Crialesi-Esposito}}, \bibinfo {author} {\bibfnamefont {G.}~\bibnamefont
  {Boffetta}}, \bibinfo {author} {\bibfnamefont {L.}~\bibnamefont {Brandt}},
  \bibinfo {author} {\bibfnamefont {S.}~\bibnamefont {Chibbaro}},\ and\
  \bibinfo {author} {\bibfnamefont {S.}~\bibnamefont {Musacchio}},\ }\bibfield
  {title} {\bibinfo {title} {Intermittency in turbulent emulsions},\ }\href
  {https://doi.org/10.1017/jfm.2023.628} {\bibfield  {journal} {\bibinfo
  {journal} {Journal of Fluid Mechanics}\ }\textbf {\bibinfo {volume} {972}},\
  \bibinfo {pages} {A37} (\bibinfo {year} {2023}{\natexlab{c}})}\BibitemShut
  {NoStop}%
\end{thebibliography}%

\end{document}